# Hemaka's constant

**Amelia Carolina Sparavigna**
Department of Applied Science and Technology
Politecnico di Torino, Italy

*As proposed in a previous paper, the decorations of ancient objects can provide some information on the approximate evaluations of constant π, the ratio of circumference to diameter. Here we discuss some disks found in the tomb of Hemaka, the chancellor of a king of the First Dynasty of Egypt, about 3000 BC. The discussion is based on measurements of the dimensionless ratio of lengths.*

Recently, I have proposed the analysis of the decorations of ancient objects to investigate the knowledge of constant π as the dimensionless ratio of circumference to diameter. In particular I have considered the decoration of the Langstrup belt plate [1]. This object is coming from Denmark, and was dated approximately 1400 BC. The analysis provided that the artist of the Bronze Age used some rational approximations of π (15/5, 22/7, 26/8 and 32/10).

It is generally believed that the estimation of π constant is quite old. According to [2] the value of π is involved in the proportions of the Great Pyramid of Giza, of the 26th century BC, in the form of 3+1/7 = 22/7 [3]. Some scholars consider this the result of a deliberate design proportion [4]. Others concluded that ancient Egyptians had no concept of π, and that the observed pyramid design was based just on the choice of its slope [2,3,5].

In ancient Egypt, the first written calculation of π can be found in the Rhind Mathematical Papyrus. The Egyptian scribe Ahmes wrote this text, which can be considered the oldest known treatise of mathematics. The Rhind Papyrus dates from the Egyptian Second Intermediate Period, but it seems that Ahmes stated that he copied a Middle Kingdom papyrus, therefore before 1650 BC [6]. In this papyrus we can find how we can approximate π. In the problem n.48 proposed by Ahmes, the area of a circle was computed by approximating the circle by an octagon. According the Ref.2, "the value of π is never mentioned or computed, however. If the Egyptians knew of π, then the corresponding approximation was 256/81." As we can see from Ref.7, the papyrus discusses the approximation of the area of circle, as shown in Fig.1. The procedure is the following. A square with the side length equal to the diameter of the circle is drawn. Then, the square (side length of 9 units) is subdivided in 9 squares, each of side length of 3 units. The area of seven such squares approximates the area of circle. Therefore π=4×7/9=28/9=3.11 (see Figure 1), having a difference from the true value of about 1%. In my opinion, this was a quite good value, considering the instruments the ancient Egyptian had to draw geometrical figures and subdivide them in equal parts [8].

Here I want to discuss an even older estimation of π, based on the decorations of some items found in the toms of chancellor Hemaka at Saqqara [9-11]. Hemaka was an important official during the reign of the First Dynasty Egyptian king Den. He was the royal seal-bearer, that is, the king's chancellor, second in power only to the king. His tomb, located in the northern part of Saqqara was enriched by many grave goods, "including an inlaid gaming disc and the earliest surviving piece of papyrus." [9]. For what concerns the period, the reign of king Den lasted from 2975 BC to 2935 BC.

Information about the first dynasty is coming from a few monuments and other objects. At that time, the tombs of kings were built of wood and mud bricks, with some small insets of stone for walls and floors. Stone was largely used for producing ornaments, vessels and statues. It seems that it was during this period that Egyptians woodworkers invented the fixed mortise and tenon joint [12]. A variation of this joint became one of the most important features in Mediterranean shipbuilding. However, we have to tell that human sacrifices were practiced during funerary rituals of the kings. This practice ended with the dynasty, and small statuettes, the shabtis, took the place of persons to serve the kings in the afterlife [13].

Hemaka is important for our research on $\pi$ because he had, as previously told some decorated disks among the items for his afterlife. Sketches in Fig.2 show the disks (these drawings have been created according to images in Ref.14 and 15, see them for the pictures). One disk (2a) is decorated in the central part with two birds, having a two-fold rotational symmetry. Disk (2b) has an interesting decoration composed of four animals, two gazelles and two jackals, in an anti-symmetric four-fold arrangement. The hunting is rendered by the rotational symmetry, as proposed in a previous paper [16]. The third disk (2c) has a geometric decoration only. The reader can imagine such disks having diameters a little bit larger than ten centimetres.

Small squares decorated disks (2a) and (2c). These decorations are really interesting for evaluating the Hemaka's knowledge of $\pi$. Let us start our discussion from (2c): at Ref.14, we can find an interesting but rather complex discussion of the decoration, based on the drawing of squares and circles. Here I propose a more practical approach, following the method already used for the Langstrup plate. I consider the constant $\pi$ as the dimensionless ratio of two measured quantities: circumference and diameter lengths. To measure these quantities I used as unit of measure the diagonal $d$ of the small white squares of the decoration. For the measures on (2c), see Figure 3 on the left. We have therefore the data given in Table I.

Table I

| Annulus | D, Diameter | C, Circumference | $\pi$=C/D |
|---|---|---|---|
| I | 5 $d$ | 17 $d$ | 17/5=34/10=3.4 |
| II | 8 $d$ | 26 $d$ | 26/8=52/16=3.25 |
| III | 11 $d$ | 33 $d$ | 33/11=66/22=3. |

In Table I we have fraction 26/8, the value that we have already seen used in the Langstrup plate. Problems concerning the arrangement of small squares in I and II annuli (annulus is the region lying between two concentric circles), arise from the fact that C is larger than the right one (there is one more square than necessary). Problems were solved by the artist with a small rotation about their centres of some squares. For the larger annulus, III, this problem does not exist. As a result of measurements, for this third annulus we have what is generally considered as the oldest approximation of $\pi$, that is, 3.

This same ratio 33/11 appears in the decoration of disk (2a). In fact, we need to digitally restore the decoration, as proposed in Figure 3 on the right. The best restoration I have obtained is that using 33 squares. If we choose a diameter of 10 times $d$, we have $\pi$=3.3; but, if we considered the diameter as (10+1/2)$d$ instead of 11 or 10$d$, we had $\pi$ evaluated as 33/(10+1/2)=66/21=3.1428. If we assume an uncertainty of about $d$/2 for the measures of diameters on (2c), we could arrange the following Table II.

Table II

| Annulus | D, Diameter | C, Circumference | $\pi$=C/D |
|---|---|---|---|
| I | (5+1/2) $d$ | 17 $d$ | 34/11=3.09 |
| II | (8+1/2) $d$ | 26 $d$ | 52/17=3.06 |
| III | (10+1/2) $d$ | 33 $d$ | 66/21=3.14 |

In my opinion, Hemaka knew that the value of π was between two fractions, as, for instance, the best approximation 66/21 we found is between 66/22 and 66/20. Therefore, the decorations were probably created using some tables based on integers and their ratios, the fractions, developed for practical purposes. These persons, that accompanied their funeral rites with human sacrifices, had probably quite good practical knowledge of mathematics.

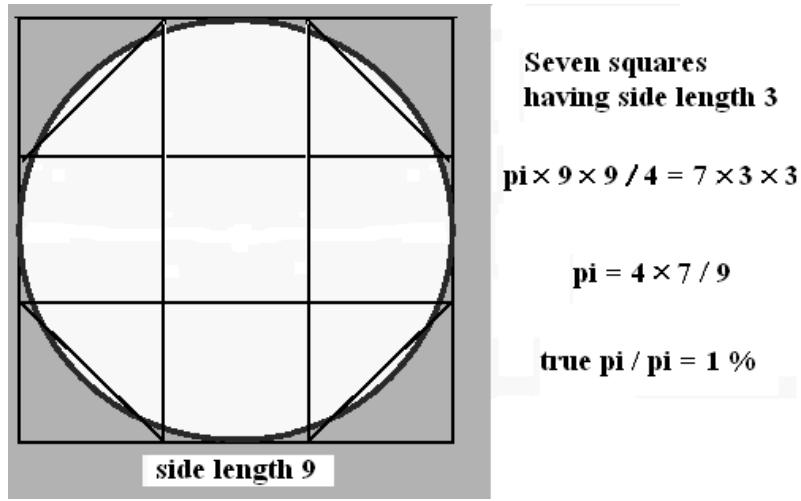

Fig.1. The figure shows how to approximate the area of the circle using squares, as proposed by scribe Ahmes in the Rhind Papyrus, Egyptian Second Intermediate Period.

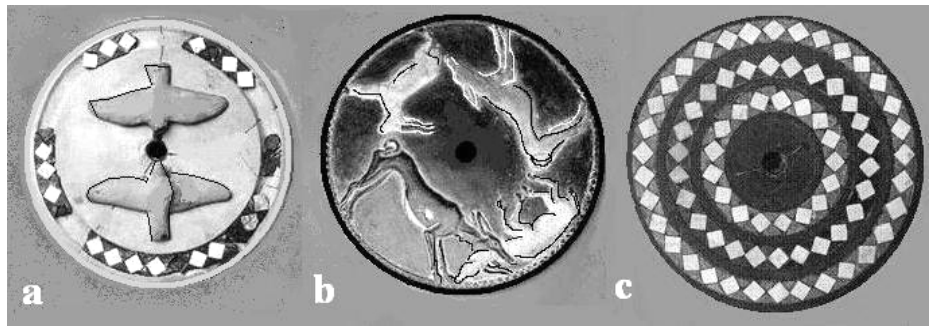

Fig.2 These sketches represent the Hemaka's gaming disks.

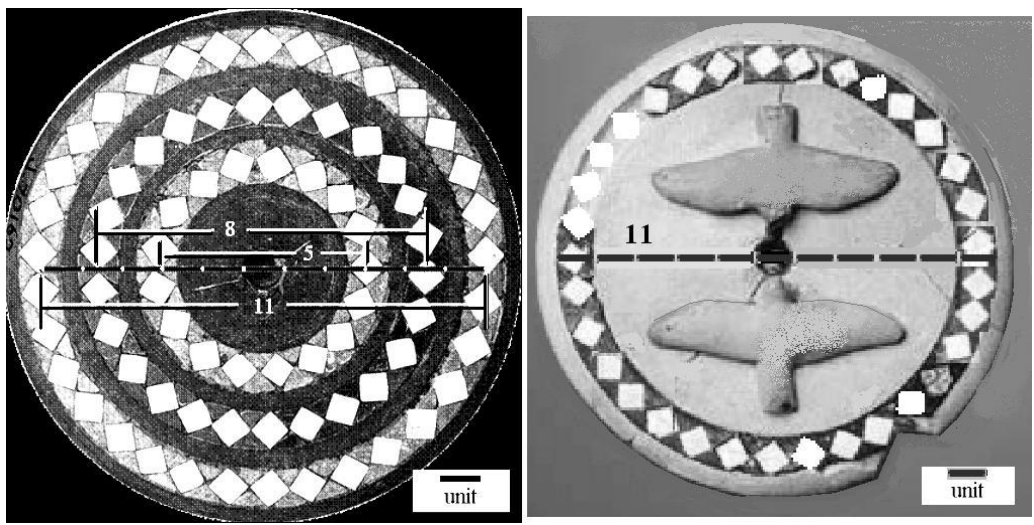

Fig.3 According to a given unit of measure (the diagonal $d$ of the small white square) we can measure circumferences and diameters, having therefore the possibility to estimate $\pi$. The disk (2a) had been subjected to a digital restoration.